\documentclass[runningheads]{llncs}
\usepackage{url}
\usepackage{graphicx}
\usepackage{amsmath,amssymb}
\usepackage{float}
\usepackage{comment}
\usepackage[colorinlistoftodos]{todonotes}
\usepackage{multicol}

\newcommand*\samethanks[1][\value{footnote}]{\footnotemark[#1]}

\begin{document}
\title{Prediction of MGMT Methylation Status of Glioblastoma using Radiomics and Latent Space Shape Features}
\titlerunning{MGMT Prediction using Radiomics and Latent Shape Features}

\author{Sveinn P\'{a}lsson\inst{1} \and
Stefano Cerri\samethanks\inst{1} \and
Koen Van Leemput\inst{1,2}}

\institute{Department of Health Technology, Technical University of Denmark, Denmark
\and
Athinoula A. Martinos Center for Biomedical Imaging, Massachusetts General Hospital, Harvard Medical School, USA}
\maketitle              
\begin{abstract}

In this paper we propose a method for predicting the status of MGMT promoter methylation in high-grade gliomas. From the available MR images, we segment the tumor using deep convolutional neural networks and extract both radiomic features and shape features learned by a variational autoencoder. We implemented a standard machine learning workflow to obtain predictions, consisting of feature selection followed by training of a random forest classification model. We trained and evaluated our method on the RSNA-ASNR-MICCAI BraTS 2021 challenge dataset and submitted our predictions to the challenge.

\keywords{ MGMT prediction \and radiomics \and deep learning \and glioblastoma \and variational autoencoder}
\end{abstract}

\section{Introduction}
\label{sec:intro}

Expression of O$^6$-methylguanine-DNA-methyltransferase (MGMT) in glioblastoma is of clinical importance as it has implications of the patient's overall survival \cite{Michaelsen2013,Gorlia2008}. The prognostic information of MGMT is believed to be due to resistance of tumors with unmethylated MGMT promoter to Temozolomide \cite{Hegi2005,Kitange2009}, a drug used in standard therapy \cite{Stupp2009}. Inference of the MGMT status in the clinic is done by histological analysis, as currently available non-invasive techniques are still too unreliable. 

The RSNA-ASNR-MICCAI BraTS 2021 challenge \cite{Baid2021,Menze2014,Bakas2018,bakas2017advancing,bakas2017lgg,bakas2017gbm} contains two tasks: tumor segmentation and MGMT methylation prediction from pre-operative magnetic resonance (MR) images. The challenge organizers have released a large dataset with the goal of facilitating comparison between methods and advancing state-of-the-art methods in these domains. In this paper we focus on the prediction task only.

Radiomics~\cite{Lambin2012} is a method for extracting features from MR images. The features, called ``radiomic'' features are a variety of statistical, shape and texture features, extracted from a target region within an MR image. Radiomics has gained much interest for prediction tasks related to brain tumors \cite{Booth2020} and has been successfully applied to MGMT methylation prediction \cite{Xi2018,Li2018}.

We propose a method for inference of the MGMT methylation that combines the use of radiomics with shape features learned by a variational autoencoder (VAE)~\cite{Kingma2013}.
VAE, implemented with deep neural networks, can learn high level features that are specific to the data structure it is trained on. By training the VAE on tumor segmentations, we may be able to extract complex tumor shape features that radiomics does not include. Combining hand-crafted features with a learned latent representation of medical images for classification has been previously studied \cite{Cui2019}, showing improved model classification performance.

The paper is structured as follows: In Section~\ref{sec:methods} we describe our methods in detail. In Section~\ref{sec:results}, we present our results and in Section~\ref{sec:discussion} we discuss our results and conclude.

\section{Methods}
\label{sec:methods}

\begin{figure}[H]
    \centering
    \includegraphics[width=\textwidth]{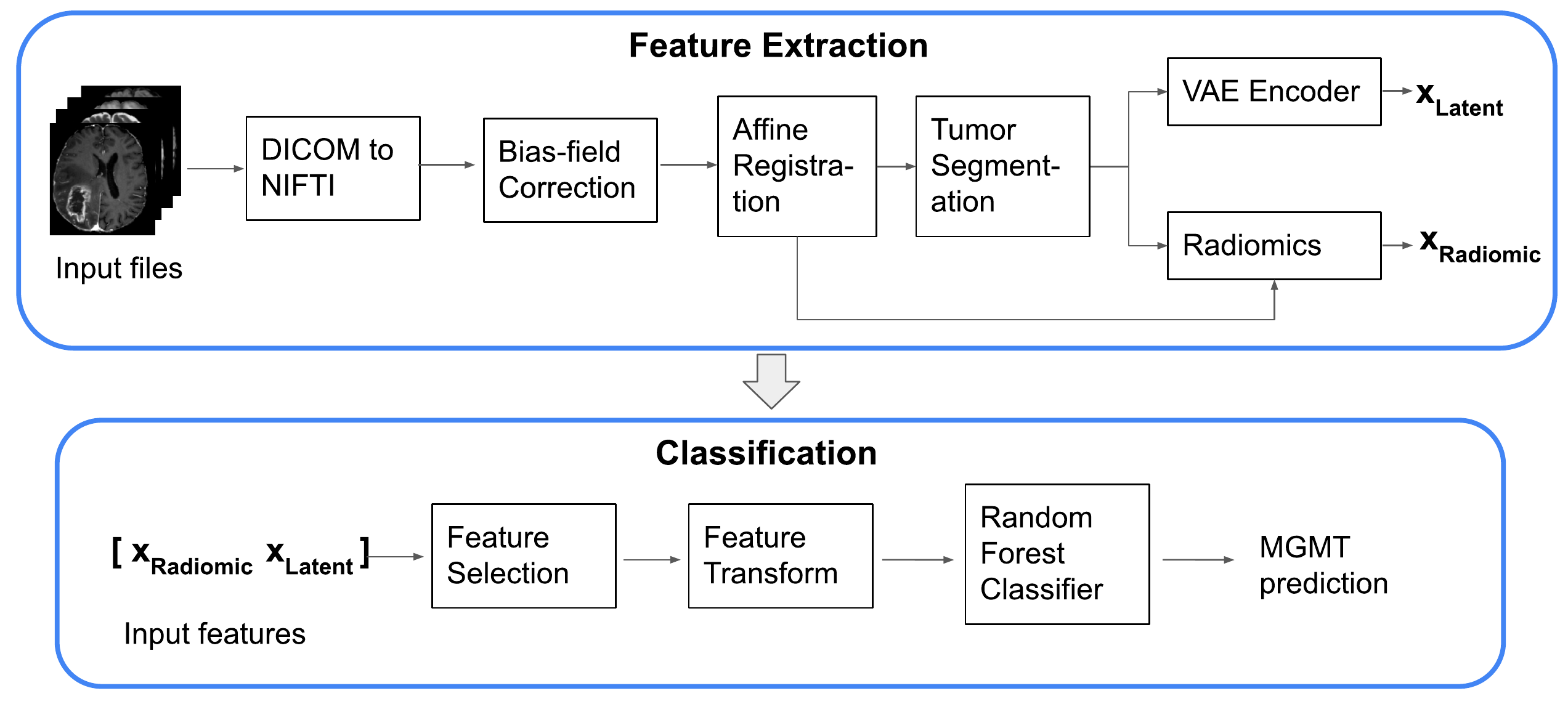}
    \caption{Overview of our method. The figure shows the main components involved in going from input images to MGMT methylation prediction.}
    \label{fig:overview}
\end{figure}

In this section we give a detailed description of our methods (illustrated in Fig.~\ref{fig:overview}). We start by describing the datasets we use. We then describe our pre-processing pipeline, consisting of DICOM to NIFTI conversion, bias-correction and registration. Next, we describe how we obtain tumor segmentations, and how we use these segmentations to compute radiomics and latent shape features. 

Finally, we describe the classification model.

\newpage

\subsection{Data}
\label{sec:data}

The challenge data consists of pre-operative MR images of 2000 subjects, divided into training, validation and testing cohorts \cite{Baid2021}. For the segmentation task, 1251 subjects are provided with ground truth labels for training segmentation models, whereas for the classification task, ground truth MGMT labels are provided for 585 of those subjects. The testing cohort is unavailable but our methods will be tested on it once it is released. Validation data for the classification task consists of image data for 87 subjects that are provided without ground truth labels but they can be used to evaluate model performance by submitting predictions to the challenge's online platform~\cite{kaggle}.

For every subject, the available modalities are T1 weighted, post-contrast T1 weighted (Gadolinium), T2 weighted and T2-FLAIR (Fig.~\ref{fig:mri_seg} (A-D)). A detailed description of the data and pre-processing applied to it by the challenge organizers is given in \cite{Baid2021}. The segmentation task dataset has been registered to a standard template and is provided as NIFTI files while the classification data are not co-registered and are provided as DICOM files.

\subsection{Pre-processing}
\label{sec:preprocessing}
Our pre-processing pipeline starts with conversion of the provided DICOM files to NIFTI (implemented in python~\cite{dicom2nifti}). Bias field correction is then performed using N4 bias field correction implemented in SimpleITK \cite{simpleITK}. We then register the T1 image to a template T1 image and subsequently register the other modalities to the newly registered T1 image. The template we use is the T1 image of a subject (id='00001') in the BraTS21 segmentation challenge. Affine registration was performed using the ANTs registration tool (implemented in python \cite{ants}).

\subsection{Tumor segmentation}
\label{sec:image_segmentation}

\begin{figure}[H]
    \centering
    \includegraphics[width=\textwidth]{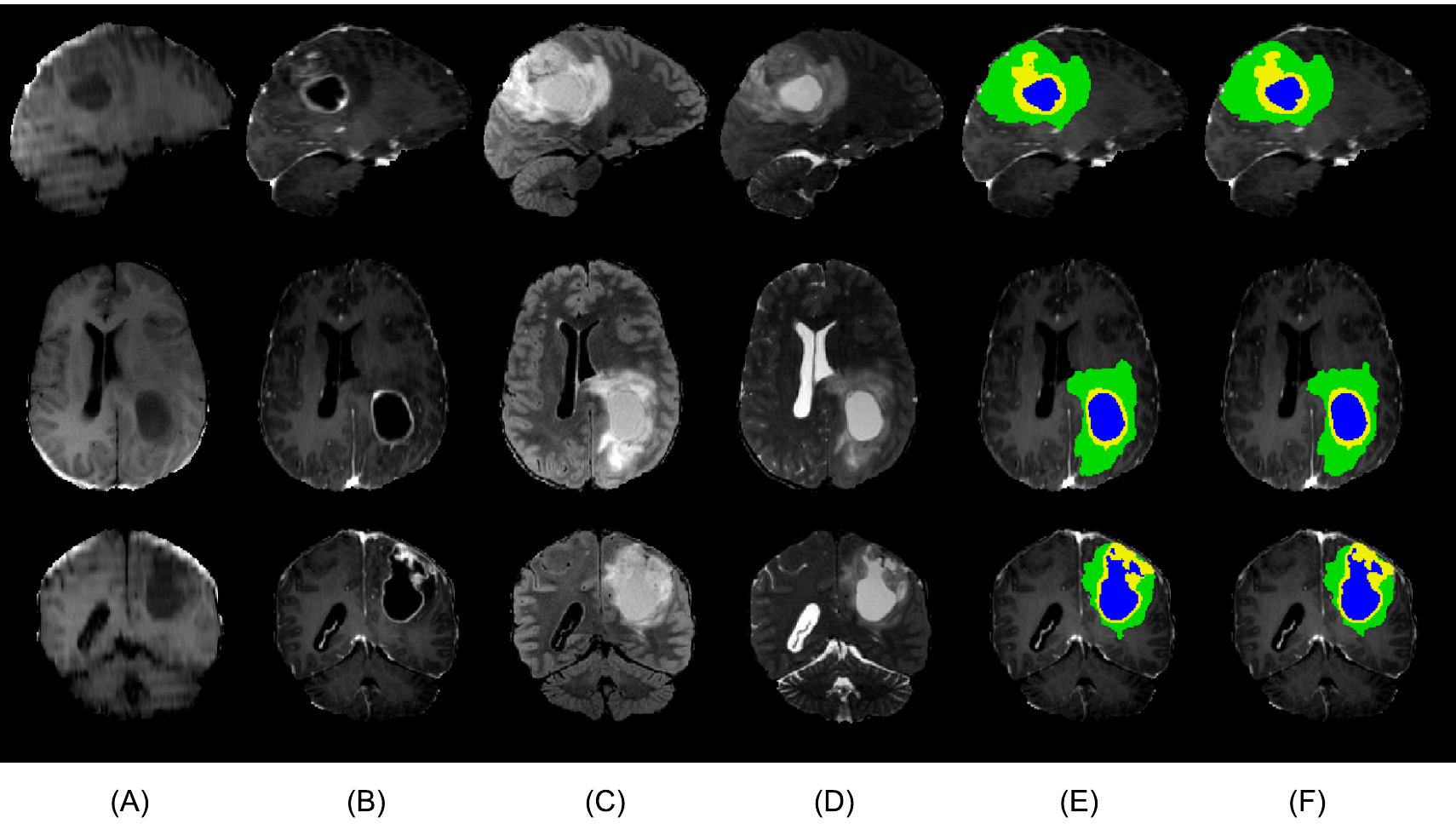}
    \caption{The figure shows an example from the challenge dataset. From top to bottom: sagittal, axial and coronal view. Columns show (A) T1w, (B) T1c, (C) FLAIR, (D) T2w, (E) ground truth tumor segmentation, (F) automatic segmentation.
    }
    \label{fig:mri_seg}
\end{figure}

As mentioned in Section~\ref{sec:data}, the tumor segmentation challenge provides 1251 images for training a segmentation model. To get accurate segmentations for further analysis, we use the winning method of the BraTS 2020 challenge \cite{Isensee2021nnu,Isensee2021}, an ensemble of deep convolutional neural networks with ``U-Net'' \cite{Ronneberger2015} style architecture, which we train on the whole set of 1251 images. The model is trained to segment three different tumor components; enhancing core, non-enhancing core and edema. The resulting model achieves high segmentation performance (a representative sample is shown in Fig.~\ref{fig:mri_seg} (F)). The resulting model is used to segment the images provided for the classification task. 

\subsection{Latent shape features}
\label{sec:shape_features}

We obtain our latent shape features from a variational autoencoder (VAE)~\cite{Kingma2013}, implemented with 3D convolutional neural networks in tensorflow \cite{tensorflow}. The VAE model consists of two networks; a decoder, designed to generate tumor segmentations from latent variables;
and encoder, to infer latent variables when given tumor segmentations. 
The input to the encoder network is a segmentation with size (240,240,155,4) where the last dimension is a one-hot encoding of the tumor component (or background) present at each voxel. The encoder network consists of 3 convolutional network blocks, followed by two fully connected layers. Each block consists of 2 convolutional layers followed by a max pooling layer. The decoder network has a symmetrical architecture to the encoder, where the convolutional layers are replaced with deconvolutional layers \cite{Dosovitskiy2015}. After each convolutional layer in both networks, a leaky ReLU \cite{Nair2010} activation is applied, except at the last layer of the decoder whose output is interpreted as logits. The VAE is trained using the ADAM optimization algorithm \cite{adam}.

We train the VAE on the 1251 available segmentations from the segmentation training dataset. To extract features from a given tumor segmentation, it is passed through the encoder network and its output is taken as the latent features. We set the number of latent features to 64.

\subsection{Radiomics}
\label{sec:radiomics_features}

We extract radiomic features from three automatically segmented tumor regions and from each provided modality, resulting in a total of 1172 extracted radiomic features. The radiomic features comprise seven categories: first-order statistics, shape descriptors, gray level co-occurrence matrix (GLCM), gray level run length matrix (GLRLM), gray level size zone matrix (GLSZM), gray level dependece matrix (GLDM), and neighboring gray tone difference matrix (NGTDM). We use the PyRadiomics \cite{Griethuysen2017} python implementation of radiomics for the feature extraction.

The three tumor regions we consider is the whole tumor, enhancing core and non-enhancing core. The whole tumor is the union of all the three tumor components that are segmented.

\subsection{Classification}
\label{sec:classifier}

After obtaining all of our features, in the next step we perform feature pre-processing to standardize feature values and feature selection to reduce dimensionality. We then train a random forest classifier on the selected features. 

For each feature, we search for a threshold value that best splits the subjects in terms of the target variable. Specifically, for each candidate threshold value, we perform a Fisher's exact test~\cite{Fisher1922}, testing the hypothesis that the binomial distributions (over the target variable) of the two resulting groups are the same. The feature value resulting in the lowest P-value is chosen as the threshold for that feature. The features are subsequently transformed to binary variables according to which side of the threshold they land. Features are then selected if the P-value of the best threshold is $P<P_{\text{min}}$, where $P_{\text{min}}$ is experimentally chosen. The selected features and choice of $P_{\text{min}}$ will be discussed further in Section~\ref{sec:feature_selection}.

We use
a random forest \cite{Breiman2001} (implemented in python~\cite{sklearn}) 
to obtain predictions of MGMT methylation status, given the input features we extracted. The model is trained on the 585 available subjects via K-fold cross validation, with K chosen such that in each fold, 5 subjects are held out while the remaining subjects are used to train a model ($K=117$ in our case). In each fold, the model is trained on 580 subjects and predictions on the 5 held-out subjects are obtained. Once predictions are obtained for all subjects, a performance score is calculated. The performance score we use is the area under the receiver operating characteristic curve (AUC). Using grid search, we tune two hyperparameters of the model; the number of samples to split a node and maximum depth of trees.
At test time, given an unseen subject, the 117 models are all used to predict the MGMT methylation status, each predicting either 0 or 1 for the unmethylated or methylated group, respectively. The average of the predictions is interpreted as the probability of belonging to the methylated group.

\section{Results}
\label{sec:results}

\subsection{Feature Selection}
\label{sec:feature_selection}

The number of features selected by the selection procedure described in Section~\ref{sec:classifier} depends on our choice of $P_{\text{min}}$, which we experimentally determine by searching a range of values and measuring model performance using the whole training cohort. We set $P_{\text{min}} = 5\times10^{-4}$ which leaves 23 features remaining; 16 of which are radiomic features and 7 latent shape features. The list of selected radiomic features is given in Table~\ref{tab:selected_features}. We observe selected radiomic features from 6 out of 7 categories mentioned in~\ref{sec:radiomics_features}, from 3 out of 4 modalities and from all 3 tumor regions. 

\begin{table}
 \caption{List of selected radiomic features.}
 \label{tab:selected_features}
\tabcolsep=0.185cm
 \begin{tabular}{llll} 
 Category & Feature name & Modality & Region\\ [0.5ex] 
 \hline
    Shape & Maximum 3D Diameter & - & Enh-core\\
    First order & Interquartile Range & T1-ce & Core\\
    First order & Mean Absolute Deviation & T1-ce & Core\\
    First order & Mean & T1-ce & Core\\
    First order & Median & T1-ce & Core\\
    First order & Median & T1-ce & Whole\\
     First order & Variance & T1-ce & Core\\
First order & 10Percentile & FLAIR & Core\\
GLRLM & Graylevel non-uniformity normalized & FLAIR & Whole\\
GLRLM & Graylevel variance & FLAIR & Whole\\
GLSZM & Small area emphasis & FLAIR & Whole\\
GLSZM & Small area high graylevel emphasis & FLAIR & Whole\\
GLSZM & Small area low graylevel emphasis & FLAIR & Whole\\
NGTDM & Busyness & FLAIR & Whole\\
GLSZM & Small area high graylevel emphasis & T2 & Whole\\

 \end{tabular}
\end{table}

\subsection{Classification}
\label{sec:clf_results}

We find the best hyperparameters for the random forest through grid search to be 2 samples to split a node and a maximum tree depth of 4 (we leave other parameters as default). The whole training cohort is used for the hyperparameter search.

To test the benefit of using the latent features in the model along with the radiomic features, we train the model on both feature sets separately and together and measure the AUC score. For a more accurate performance measure on the training dataset, we ran our cross validation 10 times (each time the dataset is shuffled) and in Table~\ref{tab:results}, we report the mean AUC score across the 10 iterations. The true labels of the validation dataset are unknown to us, but by submitting our predictions to the challenge platform, we obtain a validation AUC score reported in Table~\ref{tab:results}. We observe a substantial disagreement between the training and validation scores: the training results show improvement with the combination of feature sets, while the validation scores indicate that using radiomics alone is preferred and that the latent shape features have very low predictive value.

\begin{table}
\centering
 \caption{Classification performance measured by AUC. For three feature sets, the table shows AUC score for both cross-validated training set predictions and predictions on the validation set.}
 \label{tab:results}
\tabcolsep=0.185cm
 \begin{tabular}{l | c | c} 
 Features & Training  & Validation \\ [0.5ex] 
 \hline
  Radiomics + Latent  & 0.603  & 0.598\\
  Radiomics  & 0.582 & 0.632\\
  Latent  & 0.568 & 0.488\\

 \end{tabular}
\end{table}

\vspace{-0.8cm}

\section{Discussion}
\label{sec:discussion}

In this paper, we propose a method for MGMT methylation prediction that combines the use of radiomics with high level shape features learned by a variational autoencoder. We train a segmentation model to obtain tumor segmentations, and train a variational autoencoder on segmentations to learn high-level shape features of tumor. We use the tumor segmentation to compute radiomic features, and pass the segmentation to the encoder network of the variational autoencoder to obtain shape features from its latent space. We extracted these features from the training data provided by the RSNA-ASNR-MICCAI BraTS 2021 challenge and trained a random forest classifier. The method was submitted to the challenge and obtained a validation score (AUC) of 0.598.

As we discussed in Section~\ref{sec:intro}, radiomic features have already been shown to be applicable to this prediction task while tumor shape has not been proven to predict MGMT methylation. Therefore, to test whether the feature set combination we propose performs better than simply using the radiomic features alone, we experiment with training the classifier on them separately. On our training data, we observe a performance benefit of using the shape features (cf.~Table ~\ref{tab:results}), but this is not reproduced on the validation set where the radiomic features alone achieve a score of 0.632 but the latent features only 0.488. This may be due to overfitting of our feature transform and hyperparameter selection to the training data or high uncertainty stemming from the small number of samples in the validation dataset. We hope to gain more insight by submitting our method to the testing phase of the challenge, which contains a substantially larger number of subjects.

\section{Acknowledgements}
This project was funded by the European Union’s Horizon 2020 research and innovation program under the Marie Sklodowska-Curie project TRABIT (agreement No 765148) and NINDS (grant No R01NS112161). 

\newpage

\bibliographystyle{unsrt}
\bibliography{refs}

\end{document}